\documentclass{Interspeech}
\usepackage{xurl}
\usepackage{booktabs}
\usepackage{array}
\newcolumntype{L}[1]{>{\raggedright\arraybackslash}p{#1}}
\newcommand{\acctext}{\ensuremath{\mathrm{Score}_{\mathrm{text}}}}% text-input accuracy
\newcommand{\accspeech}{\ensuremath{\mathrm{Score}_{\mathrm{speech}}}}% speech-input accuracy
\usepackage{graphicx}
\providecommand{\tightlist}{}
\usepackage{fontawesome5}

\makeatletter
\newcommand\blfootnote[1]{%
  \begingroup
  \renewcommand\thefootnote{}%
  \long\def\@makefntext##1{\parindent=0pt\noindent##1}%
  \footnote{#1}%
  \addtocounter{footnote}{-1}%
  \endgroup
}
\makeatother

\title{SEA-HELM Audio: Evaluating Multilingual ASR-LLM Pipeline Performance Across SEA-HELM Subsets}

\author{Aulia}{Adila}
\author{Yixian}{Tan}
\author{Jonathan}{Heng}
\author{Kunat}{Pipatanakul}
\author{Donghang}{Wu}
\author{Anocha\nobreak}{Sutaveephamochanon}

\makeatletter
\def\affiliationslist{AI Singapore, Singapore}% one shared affiliation, no superscripts
\makeatother
\email{aulia.adila,yixian,jonathan,kunat.p,donghang,anocha@aisingapore.org}

\begin{document}
\maketitle
% \blfootnote{% \\
%   ~Leaderboard: \url{https://voice-agent.sea-lion.ai/leaderboard/}\\
%   ~Dataset: \url{https://huggingface.co/datasets/aisingapore/seahelm-audio}\\
%   ~Code: \url{https://github.com/aisingapore/seahelm-audio}
% }
\blfootnote{\scriptsize
  \begin{tabular}{@{}l@{~}l@{}}
    ~Leaderboard: \url{voice-agent.sea-lion.ai/leaderboard} \\
    ~Dataset: \url{huggingface.co/datasets/aisingapore/seahelm-audio} \\
    ~Code: \url{github.com/aisingapore/seahelm-audio} \\
  \end{tabular}}

\begin{abstract}
We introduce \textbf{SEA-HELM Audio}, a Southeast Asian (SEA) multilingual evaluation framework that extends the text-based SEA-HELM benchmark to assess performance degradation in cascaded automatic speech recognition and large language model (ASR–LLM) pipelines. Using synthesized speech, we compare speech-input and text-input performance across five languages—Chinese, English, Malay, Indonesian, and Tamil—on multiple-choice reasoning and instruction-following tasks. We evaluate open-weight and commercial ASR–LLM systems using identical verifiers and quantify the \textbf{speech gap} as the difference between text and speech accuracy. Results show substantial degradation, with accuracy drops of 9.00–18.50 percentage points for multiple-choice reasoning and 20.00–42.86 points for instruction following. Tamil exhibits the largest degradation, exceeding 40 percentage points. Our findings highlight the limitations of text-only benchmarks in predicting voice-agent reliability and demonstrate the value of SEA-HELM Audio for evaluating multilingual voice-agent systems.
\end{abstract}

\section{Introduction}
\label{introduction}

As artificial intelligence systems transition from text-based interfaces to conversational voice-enabled agents, evaluating the end-to-end accuracy of cascaded speech-language stacks has become a critical research priority. While text-based Large Language Models (LLMs) continue to achieve high benchmark scores, deploying these models within voice pipelines introduces acoustic transcription errors, phonetic ambiguities, and domain-specific degradation. Quantifying this degradation, specifically measuring the accuracy penalty a user incurs by speaking rather than typing, is essential for understanding real-world agent reliability across diverse tasks.

Existing spoken language model benchmarks, such as VoiceBench~\cite{chen2024voicebench}, evaluate capabilities including speech recognition, reasoning, and instruction following through spoken inputs. However, their predominantly English-language evaluation settings provide limited insight into how voice-agent systems perform on Southeast Asian languages and tasks. In particular, it remains unclear to what extent transcription errors alter task-critical information and how these errors affect downstream task performance, especially in the context of Southeast Asian languages.

To address the need for localized evaluation, AI Singapore developed \textbf{SEA-HELM} (Southeast Asian Holistic Evaluation of Language Models)~\cite{susanto2025seahelm}. SEA-HELM provides authentic per-language prompts and tailored programmatic verifiers across regional languages. However, SEA-HELM was engineered natively as a text-only benchmark, consisting exclusively of written prompts without paired audio recordings.

This paper addresses the central research question: \textbf{How can a text-based, regionally grounded benchmark be adapted into a speech-based evaluation dataset to systematically measure the impact of speech input on downstream LLM task performance across Southeast Asian languages?} By bridging SEA-HELM's text prompts with synthesized audio pipelines, we establish a framework to quantify the \textbf{Speech gap}, the difference between \acctext{} and \accspeech{}, across multiple Southeast Asian languages, providing empirical baselines for regional speech agent deployment.

\section{Methodology}
\label{methodology}

This section outlines the experimental framework, dataset selection, and evaluation metrics employed to systematically isolate and quantify the performance degradation introduced by the speech modality across our five target languages.

\subsection{Experimental Design}
\label{experimental-design-and-rationale}

Prior to finalizing the evaluation pipeline, we conducted an exploratory phase to establish a robust and regionally relevant testing environment. A primary distinction between our setup and existing voice evaluation frameworks lies in audio corpus generation: VoiceBench uses pre-recorded human speech, whereas SEA-HELM provides only text prompts. Consequently, our cascade architecture synthesizes its own input speech before evaluating downstream models.

To ensure audio fidelity across regional languages, we generated synthetic speech with OmniVoice~\cite{zhu2026omnivoice} Text-to-Speech (TTS) engines, selected for acoustic naturalness, prosodic accuracy, and regional accent representation.

For task design, we selected multiple-choice and instruction-following formats, which are established in voice evaluation, and instantiated them with SEA-HELM text prompts. This approach preserves regional nuances, code-switching patterns, and local cultural contexts across Chinese, Malay, Tamil, Indonesian, and English. By evaluating each sample twice, once through the synthesized voice pipeline (\accspeech{}) and once via direct text prompts (\acctext{}), we establish a controlled baseline to isolate and measure the precise performance penalty imposed by the speech-to-text layer.

\subsection{Datasets and Task Selection}
\label{datasets-and-task-selection}

To evaluate the voice cascade cost-effectively and deterministically, we selected tasks from SEA-HELM that rely on \textbf{rule-based scoring} and closely align with established VoiceBench task types.

The evaluation suite draws from four primary SEA-HELM task subsets, summarized in Table~\ref{tab:subsets}.

\begin{table*}[t]
\centering
\caption{SEA-HELM tasks considered for SEA-HELM Audio and their evaluation scope.}
\label{tab:subsets}
\scriptsize
\setlength{\tabcolsep}{3pt}
\begin{tabular}{L{0.15\textwidth}L{0.12\textwidth}L{0.07\textwidth}L{0.17\textwidth}L{0.40\textwidth}}
\toprule
\textbf{Task} & \textbf{Task Type} & \textbf{SEA-HELM Languages} & \textbf{Evaluated Here} & \textbf{Selection Rationale} \\
\midrule
SEA-HELM IFEval & Instruction following~\cite{zhou2023ifeval} & 10 & English, Malay, Indonesian, Tamil & Prompts are plain instructions that can be spoken as written. Scored by SEA-HELM's multilingual verifier, which also verifies the response language. \\
Global-MMLU-Lite & Multiple choice~\cite{singh2025globalmmlu} & 7 & Chinese, English & Short questions with four options; the answer is a single letter, so scoring is exact. \\
SEA-HELM NLU QA & Question answering & 5 & Future work & Scored with SEA-HELM's rule-based token-F1 matching. \\
SEA-HELM Safeguard & Harmful-request refusal & 8 & Future work & Harmful prompts used as a refusal probe (analogous to AdvBench~\cite{zou2023advbench} in VoiceBench), testing whether the LLM still refuses when the request is spoken. \\
\bottomrule
\end{tabular}
\end{table*}

\subsubsection{Benchmark Alignment and Active Subset Scope}
\label{benchmark-alignment-and-active-subset-scope}

For the initial empirical runs detailed in this study, benchmarking focuses on the SEA-HELM IFEval and Global-MMLU-Lite subsets across our five focus languages (Chinese, Malay, Tamil, Indonesian, and English).

The selection criteria prioritized two key attributes:

\begin{enumerate}
\def\labelenumi{\arabic{enumi}.}
\tightlist
\item
  \textbf{Speakable Prompts:} Prompts can be read aloud as natural speech without removing formatting or template markup first.
\item
  \textbf{Verifiable Evaluation:} All scoring is executed via programmatic, rule-based verifiers, so the text path and the speech path are scored in exactly the same way.
\end{enumerate}

\subsection{Evaluation Metrics}
\label{evaluation-metrics}

To formalize the measurement of the speech modality penalty, the following metrics were captured across every benchmark run:

\begin{itemize}
\tightlist
\item
   \textbf{Text accuracy (\acctext{}):} The accuracy of the target LLM when provided with the original, ground-truth text input. This represents the theoretical performance ceiling for the task.
\item
  \textbf{Speech accuracy (\accspeech{}):} The end-to-end accuracy of the target LLM when the prompt is spoken, synthesized by TTS and transcribed by ASR.
\item
  $\Delta$\textbf{ (The Gap):} $\Delta = \acctext - \accspeech$. This absolute difference represents the modality penalty, quantifying how much task performance degrades due to transcription errors, phonetic ambiguities, or missing acoustic signals.
\end{itemize}

\section{SEA-HELM Audio}
\label{seahelm-audio}

SEA-HELM contains only written prompts, so SEA-HELM Audio creates the spoken input itself: each prompt is read aloud by a speech synthesizer, transcribed by the ASR system under test, and answered by the LLM. Comparing this against the same LLM answering the original text isolates the cost of the speech front-end.

\subsection{Evaluation Flow}
\label{evaluation-flow}

\begin{figure*}[h!]
\centering
\includegraphics[width=0.85\textwidth]{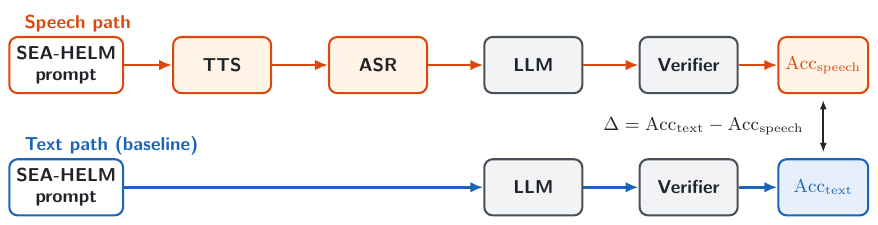}
\caption{SEA-HELM Audio evaluation: speech path (TTS $\rightarrow$ Speech $\rightarrow$ LLM) vs.~text path (LLM only).}
\label{fig:flow}
\end{figure*}

Each SEA-HELM prompt is evaluated in two ways (Figure~\ref{fig:flow}). In the \textbf{speech path}, the prompt is converted to speech with TTS, transcribed back to text by ASR, and the transcript is answered by the LLM. In the \textbf{text path}, the LLM answers the original prompt directly. Both answers are scored by the same rule-based verifier, giving \accspeech{} and \acctext{}. Because the two paths share the same LLM and the same verifier, the difference between them ($\Delta$ = \acctext{} $-$ \accspeech{}) is the accuracy lost by speaking the prompt instead of typing it.

\subsection{Spoken Prompt Construction}
\label{spoken-prompt-construction}

The TTS system plays the role of the user. Its voice is chosen independently of the system being evaluated, so every ASR--LLM configuration hears the same spoken prompts and differences in \accspeech{} reflect the system rather than the input.

Only the content a user would actually say is synthesized. For multiple-choice tasks, this is a short instruction followed by the question and its four options; for instruction-following tasks, it is the prompt exactly as written. Written conventions that do not read naturally aloud are adapted per language: in Chinese, the option labels A--D and the answer marker are written as Chinese words. The answer key (the correct option, or the constraints an IFEval response must satisfy) is never spoken and is used only at scoring time.

\subsection{Speech Recognition}
\label{speech-recognition}

The ASR system turns the synthesized speech back into text. We evaluate three systems chosen for their coverage of the target languages (Table~\ref{tab:asr_settings}): Qwen3-ASR-1.7B~\cite{shi2026qwen3asr} as an open-weight multilingual baseline, MERaLiON3-ASR-3B~\cite{he2024meralion} as a regionally adapted open-weight model for Tamil, and ElevenLabs Scribe Realtime as a commercial service. Open-weight models use greedy decoding so that transcripts are reproducible.

Transcription quality is also measured directly, as Word Error Rate (WER) and Character Error Rate (CER) between the transcript and the original prompt. This separates how much the ASR changed the prompt from how much those changes affected the task.

\begin{table}[t]
\centering
\caption{ASR systems evaluated.}
\label{tab:asr_settings}
\scriptsize
\setlength{\tabcolsep}{3pt}
\begin{tabular}{L{0.29\columnwidth}L{0.19\columnwidth}L{0.22\columnwidth}L{0.18\columnwidth}}
\toprule
\textbf{ASR System} & \textbf{Type} & \textbf{Languages} & \textbf{Decoding} \\
\midrule
Qwen3-ASR-1.7B & Open-weight & en, zh, ms, id & Greedy \\
MERaLiON3-ASR-3B & Open-weight & ta & Greedy \\
ElevenLabs Scribe Realtime & Commercial API & en, zh, ms, id, ta & Provider default \\
\bottomrule
\end{tabular}
\end{table}

\subsection{Response Generation}
\label{response-generation}

The LLM answers the ASR transcript (speech path) and the original prompt (text path). We evaluate SEA-LION-4.5-27B~\cite{ng2025sealion}, a regional open-weight model, and Gemini 3.5 Flash, a commercial model. Both use deterministic decoding (temperature 0), so any difference between the two paths comes from the input rather than from sampling.

\subsection{Scoring}
\label{scoring}

All responses are scored with rule-based verifiers taken from SEA-HELM, which keeps scoring deterministic, cheap to repeat, and identical across the two paths. For multiple-choice questions, the chosen option letter (A--D) is extracted from the response and compared with the answer key. For instruction following, each response is checked against the constraints stated in its prompt and is counted as correct only if it also uses the expected language.

\section{Results}
\label{results}

This section presents the empirical findings from our multi-lingual voice cascade evaluations. Performance is reported across two primary axes: task type (multiple-choice questions via Global-MMLU-Lite vs.~instruction following via SEA-HELM IFEval) and pipeline configuration (Regional Open-Weight Baseline vs.~Commercial Stack).

\subsection{Multiple Choice Reasoning (Global-MMLU-Lite)}
\label{multiple-choice-reasoning-global-mmlu-lite}

Evaluations on the SEA-HELM Global-MMLU-Lite subset were conducted for Chinese and English, assessing general domain knowledge under a multiple-choice format (400 samples per run). Across both languages, the commercial stack (ElevenLabs Scribe Realtime paired with Gemini 3.5 Flash) outperformed the regional open-weight baseline in both text accuracy and speech accuracy.

However, Chinese demonstrated a distinct resilience to the modality shift compared to English. For Chinese, $\Delta$ remained between 9.00 and 10.75 points, whereas English incurred a much steeper penalty of 15.50 to 18.50 points.

\begin{table}[t]
\centering
\caption{Global-MMLU-Lite results (400 samples per run; TTS: OmniVoice). Accuracy (Speech, Text) and $\Delta$ = Text $-$ Speech, in percentage points.}
\label{tab:mcq}
\scriptsize
\setlength{\tabcolsep}{2.5pt}
\begin{tabular}{L{0.17\columnwidth}L{0.24\columnwidth}L{0.19\columnwidth}rrr}
\toprule
\textbf{Language} & \textbf{ASR Model} & \textbf{LLM} & \textbf{Speech} & \textbf{Text} & \textbf{$\Delta$} \\
\midrule
\textbf{Chinese} & Qwen3-ASR-1.7B & SEA-LION-4.5-27B & 71.50 & 80.50 & 9.00 \\
\textbf{Chinese} & ElevenLabs Scribe Realtime & Gemini 3.5 Flash & 79.75 & 90.50 & 10.75 \\

\textbf{English} & Qwen3-ASR-1.7B & SEA-LION-4.5-27B & 69.00 & 87.50 & 18.50 \\
\textbf{English} & ElevenLabs Scribe Realtime & Gemini 3.5 Flash & 76.75 & 92.25 & 15.50 \\
\bottomrule
\end{tabular}
\end{table}

\subsection{Instruction Following (SEA-HELM IFEval)}
\label{instruction-following-seahelm-ifeval}

SEA-HELM IFEval provided multiple test of deterministic instruction adherence across English, Malay, Indonesian, and Tamil (105 samples per run). The text-only baselines achieved exceptionally high scores, peaking at 97.14\% for English, but introducing the speech modality resulted in severe performance degradation across all tested languages.

The modality gap ($\Delta$) expanded significantly in instruction-following tasks compared to the MCQ evaluations. Tamil suffered the steepest performance drops (exceeding 40 points) for both the specialized MERaLiON3-ASR-3B pipeline and the commercial ElevenLabs equivalent, indicating severe structural vulnerabilities in Tamil ASR transcriptions.

While the commercial stack generally outperformed the baseline, SEA-LION-4.5-27B proved highly competitive in text accuracy for Southeast Asian languages, notably surpassing Gemini 3.5 Flash in Indonesian IFEval (93.33\% vs 89.52\%). Despite this strong text foundation, the open-weight pipeline struggled to recover from ASR-induced errors, reflecting $\Delta$ of over 25 points in Malay and Indonesian.

\begin{table}[t]
\centering
\caption{SEA-HELM IFEval results (105 samples per run; TTS: ElevenLabs v3). Accuracy (Speech, Text) and $\Delta$ = Text $-$ Speech, in percentage points.}
\label{tab:ifeval}
\scriptsize
\setlength{\tabcolsep}{2.5pt}
\begin{tabular}{L{0.17\columnwidth}L{0.24\columnwidth}L{0.19\columnwidth}rrr}
\toprule
\textbf{Language} & \textbf{ASR Model} & \textbf{LLM} & \textbf{Speech} & \textbf{Text} & \textbf{$\Delta$} \\
\midrule
English & Qwen3-ASR-1.7B & SEA-LION-4.5-27B & 69.52 & 97.14 & 27.62 \\
English & ElevenLabs Scribe Realtime & Gemini 3.5 Flash & 76.19 & 97.14 & 20.95 \\

Indonesian & Qwen3-ASR-1.7B & SEA-LION-4.5-27B & 60.95 & 94.29 & 33.33 \\
Indonesian & ElevenLabs Scribe Realtime & Gemini 3.5 Flash & 58.10 & 89.52 & 31.43 \\

Malay & Qwen3-ASR-1.7B & SEA-LION-4.5-27B & 52.38 & 84.76 & 32.38 \\
Malay & ElevenLabs Scribe Realtime & Gemini 3.5 Flash & 60.00 & 80.00 & 20.00 \\

Tamil & MERaLiON3-ASR-3B & SEA-LION-4.5-27B & 47.62 & 87.62 & 40.00 \\
Tamil & ElevenLabs Scribe Realtime & Gemini 3.5 Flash & 41.90 & 81.90 & 40.00 \\
\bottomrule
\end{tabular}
\end{table}

\section{Discussion \& Analysis}
\label{discussion-analysis}

The empirical results underscore a critical reality in voice-agent deployment: strong text-based LLM performance does not guarantee equivalent voice capabilities. The modality gap ($\Delta$) observed across our evaluations highlights significant disparities driven by language, task structure, and pipeline composition.

\subsection{Linguistic Resilience vs.~Vulnerability}
\label{linguistic-resilience-vs.-vulnerability}

The most striking finding is the sheer variance in the gap across the five evaluated languages.

\begin{itemize}
\tightlist
\item
  \textbf{Resilience in Chinese:} Chinese exhibited the lowest performance degradation across all configurations, maintaining a Delta of approximately 9.00 to 10.75 points. This suggests two compounding advantages: first, modern ASR systems (both Qwen and ElevenLabs) possess highly robust acoustic models for Chinese; second, the downstream LLMs demonstrate strong semantic recovery, able to infer correct answers even in the presence of minor phonetic transcription errors.
\item
  \textbf{The Tamil Bottleneck:} Conversely, Tamil suffered the most severe modality penalty, shedding over 40 points in accuracy when transitioning from text to voice. Despite SEA-LION-4.5-27B achieving a respectable 85.71\% baseline accuracy on perfect text, the end-to-end performance collapsed to 42.86\% utilizing the specialized MERaLiON3-ASR-3B model. This indicates a severe bottleneck at the acoustic processing layer, highlighting a critical data scarcity issue in Tamil ASR training corpora compared to high-resource languages.
\end{itemize}

\subsection{Task Format Sensitivity: MCQ vs.~Instruction Following}
\label{task-format-sensitivity-mcq-vs.-instruction-following}

The data reveals that the modality gap is heavily influenced by the nature of the evaluation task.

On Global-MMLU-Lite (MCQ), LLMs could often recover from noisy transcripts because multiple-choice reasoning primarily requires identifying key semantic concepts to select an option (A, B, C, or D). If an ASR model slightly misheard a noun, the LLM could still deduce the correct multiple-choice letter from contextual clues.

IFEval, in contrast, scores each response against explicit lexical, length, and format constraints stated in the prompt. When ASR drops or substitutes a word that carries one of these constraints, the LLM answers a different instruction, and the rule-based verifier marks the response as a failure even if it is otherwise correct. This mechanism accounts for the larger IFEval gaps of 20 to 43 points.

\subsection{Commercial vs. Open-Weight}
\label{commercial-dominance-vs.-regional-open-weight-promise}

While the commercial stack (ElevenLabs Scribe Realtime + Gemini 3.5 Flash) consistently delivered higher speech accuracy, analyzing text accuracy reveals a nuanced picture of regional model capabilities.

The SEA-LION-4.5-27B model proved highly competitive, and occasionally superior, in pure-text regional evaluations. Most notably, it outperformed Gemini in the Indonesian IFEval text baseline (94.29\% vs.~89.52\%). The fact that SEA-LION's end-to-end voice performance drops so sharply (a 33.33 point penalty) confirms that the limiting factor in the open-weight pipeline is not the LLM's regional reasoning capability, but rather the fragility of the open-weight ASR models (Qwen3-ASR-1.7B and MERaLiON3-ASR-3B) handling Southeast Asian accents and code-switching.

For organizations deploying regional voice agents, these findings suggest that heavily investing in localized ASR fine-tuning will yield a much higher return on investment than further scaling the LLM parameters.

\section{Conclusion and Future Work}
\label{conclusion-and-future-work}

This study shows that a text-native benchmark, such as SEA-HELM, can be adapted to benchmark end-to-end voice pipelines, provided the evaluation tasks are highly deterministic. By measuring the modality gap ($\Delta$) between pure text accuracy and speech accuracy across five languages, we have isolated the precise performance penalty introduced when a user speaks to a conversational agent rather than typing.

Our findings demonstrate that voice pipeline reliability is heavily dependent on the target language and the structural rigidity of the task. While Chinese multiple-choice reasoning exhibited a robust resilience to the modality shift ($\Delta < 11$ points), instruction following in Tamil suffered catastrophic degradation ($\Delta > 40$ points). Furthermore, while the commercial stack (ElevenLabs Scribe Realtime + Gemini 3.5 Flash) provided the most stable end-to-end performance, the strong pure-text baseline of SEA-LION-4.5-27B suggests that the primary bottleneck for regional open-weight deployments lies in the acoustic transcription layer, not the language model's reasoning capabilities.

\bibliographystyle{IEEEtran}
\bibliography{refs}
\end{document}